# Electrostrictive and electrostatic responses in contact mode voltage modulated Scanning Probe Microscopies

Eugene A. Eliseev[1], Anna N. Morozovska[2*], Anton V. Ievlev[3], Nina Balke[3], Peter Maksymovych[3], Alexander Tselev[3], and Sergei V. Kalinin[3†]

[1]Institute for Problems of Materials Science, NAS of Ukraine, Krjijanovskogo 3, 03142 Kiev, Ukraine

[2]Institute of Physics, NAS of Ukraine, 46, pr. Nauki, 03028 Kiev, Ukraine

[3]The Center for Nanophase Materials Sciences, Oak Ridge National Laboratory, Oak Ridge, TN 37922

Electromechanical response of solids underpins image formation mechanism of several scanning probe microscopy techniques including the piezoresponse force microscopy (PFM) and electrochemical strain microscopy (ESM). While the theory of linear piezoelectric and ionic responses are well developed, the contributions of quadratic effects including electrostriction and capacitive tip-surface forces to measured signal remain poorly understood. Here we analyze the electrostrictive and capacitive contributions to the PFM and ESM signals and discuss the implications of the dielectric tip-surface gap on these interactions.

[*] Corresponding author, anna.n.morozovska@gmail.com
[†] Corresponding author, sergei2@ornl.gov

Local electromechanical response of solids on the action of periodic electric bias applied to a tip underpins image formation mechanisms in several Scanning Probe Microscopy (SPM) techniques. In Piezoresponse Force Microscopy (PFM),[1-3] the electromechanical activity is directly related to the local polarization and hence can be used to map ferroelectric domain structures. In Electrochemical Strain Microscopy (ESM),[4-6] the response originates from the ionic motion and electrochemical reactions under the probe associated with the changes of chemical pressure and electrostriction-induced responses[7-9]. Both PFM and ESM allow for a broad spectrum of spectroscopic techniques in which response is measured as a function of time and bias, providing information about polarization switching and electrochemical activity respectively.[10-15] The classical example of such measurements is PFM voltage spectroscopy, yielding local hysteresis loops in single point or mapping modes[10, 16, 17].

In the last several years, PFM/ESM responses including local position dependent electromechanical activity, tip induced remnant charge states, and hysteresis loops, were reported to a broad variety of non-ferroelectric materials including manganites, $TiO_2$, and $LaAlO_3$-$SrTiO_3$[9, 18, 19]. Interestingly, qualitatively similar responses were observed also on the ultra-thin ferroelectric films[20-23]. These observation can be interpreted both as an evidence of ferroelecticity in these materials, or tip-induced electrochemical reactions in the bulk (similar to memristors) that maintain local electroneutrality, or as an evidence of electrostatic Coulombic forces mediated by hysteretic surface charging and/or bulk charge injection.

Correspondingly, of interest to interpretation of the PFM and ESM data, especially on-field hysteresis loops, is the mechanism of electromechanical interactions in the tip-surface junction mediated by electrostriction and electrostatic forces. Here, we introduce the concept of internal and external response, as illustrated in **Figure 1.** External response originates from electrostatic (and electorcapilary for liquid meniscus) forces acting in the tip-surface junction that act against the spring defined by contact stiffness. The internal response is induced by the mechanical response of material due to the field in the material created by the tip, and includes piezoelectric, ionic, and electrostrictive responses. Note that these definitions are basic and do not describe e.g. the hysteretic responses of material, as will be discussed below.

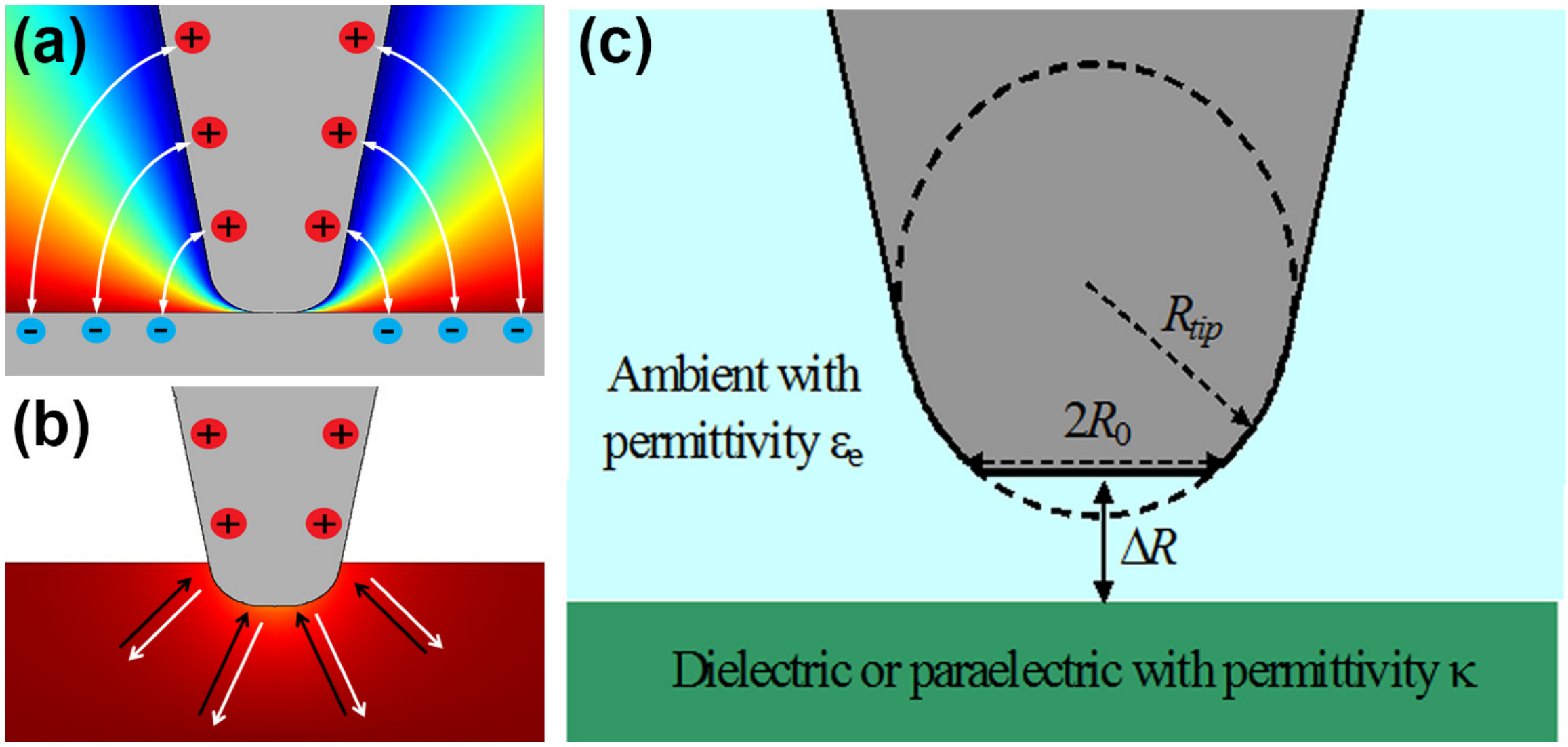

**Figure 1.** Schematics of the **(a)** external and **(b)** internal interaction between tip and surface. **(c)** Tip-surface junction model used in calculations

In general case, measured linear response can be represented as $R = a + b \cdot U_{tip}$. Here, $U_{tip}$ is the tip bias, constant $a$ comprises contributions from piezoelectric/ionic responses and static potential offset, i.e. for linear piezoelectric $R = d_{33}^{eff} + b \cdot \left(U_{tip} - U_S\right)$, where $d_{33}^{eff}$ is the effective piezoelectric tensor coefficient, $U_S$ is the surface potential. Note that the terms $d_{33}^{eff} - b \cdot U_S$ are fundamentally inseparable for local (i.e. acting on the tip) response, as analyzed in detail in Ref. [24]. In other words, the static surface potential due to surface dipoles (work function) and additional Coulombic forces due to uncompensated charge cannot be differentiated from piezoelectric response based on the bias behavior. Similarly of interest is the origin of the response slope *b*, which combines local and non-local capacitive and electrostrictive contributions and has dimension m/V. In general case,

$$b = \frac{1}{k_1 + k}\frac{dC_{tip}}{dz} + \frac{1}{k_1 + k}B + \frac{1}{24k}\frac{dC_{cant}}{dz} \tag{1}$$

where $C_{tip}$ and $C_{cant}$ are capacitances of the tip and cantilever respectively, $k$ is the cantilever spring constant, $k_1$ is the contact stiffness of the tip-surface junction, $B$ is the electrostriction constant.

Typical value of cantilever spring constant $k$ is $0.1 - 10$ N/m that is much smaller then tip-surface junction spring constant $k_1 = 10^2 - 10^3$ N/m. Thus $k$ can be neglected in electrostatic and electrostrictive contributions. As a result the magnitude of the electrostatic and electrostrictive contributions $\frac{1}{k_1}\frac{dC_{tip}}{dz}+\frac{1}{k_1}B$ is defined by contact stiffness in the tip-surface junction $k_1$. Non-local contribution $\frac{1}{24k}\frac{dC_{cant}}{dz}$ is caused by direct electrostatic interaction between cantilever and surface; it is defined by cantilever spring constant *k*. Note that while in spectroscopic modes the signal cannot be unambiguously differentiated, the imaging modes allow to attribute contrast to local or non-local based on the spatial variability of response.

In SPM, the application of the electric bias to the probe induces inhomogeneous electric field **E**, and hence polarization of the sample below the tip. For paraelectrics and linear dielectrics polarization $P_i$ and susceptibility $\chi_{ij}$ are given by expressions: $P_i \approx \chi_{ij}E_j$, $\chi_{ij} = \varepsilon_0(\varepsilon_{ij} - \varepsilon_\infty\delta_{ij})$, where $\varepsilon_0$ is the universal dielectric constant, $\varepsilon_{ij}$ is the static dielectric permittivity tensor, $\varepsilon_\infty$ is the high frequency dielectric permittivity. The electrostrictive response can be evaluated similarly to decoupled approximation[25, 26] for piezoresponse. Namely, the surface displacement induced by the SPM probe is:[27]

$$u_i^{ES+MT}(0) = -\iiint\limits_{0<\xi_3<h} \frac{\partial G_{ij}^S(-\xi_1,-\xi_2,0,\xi_3)}{\partial \xi_m} Q_{mjkl}^{MT} E_k E_l \, d^3\xi \qquad (2)$$

Here the Green's tensor $G_{ij}^S$ is given in Supplemental material, *h* is the film thickness. Electrostriction strain tensor $Q_{mjkl}^{MT}$ renormalized by the Maxwell stress is

$$Q_{ijkl}^{MT} = q_{ijkl}\chi_{pk}\chi_{ql} + \frac{1}{2}\left(\delta_{jp}\delta_{li} + \delta_{ip}\delta_{jl} - \delta_{ij}\delta_{pl}\right)\varepsilon_0\varepsilon_{pk} \qquad (3)$$

where $q_{mjkl}$ is electrostiction stress tensor.

Electric field is related with the electrostatic potential as, $E_k = -\partial\varphi/\partial\xi_k$. The potential induced by the probe is approximated using point charge model[28, 29]:

$$\varphi = \frac{Ud}{\left(\xi_1^2 + \xi_2^2 + (d + \xi_3/\gamma)^2\right)^{1/2}} \qquad (4)$$

Here *d* is the distance between the effective charge position and the sample surface, *U* is the voltage applied to the tip, $\gamma = \sqrt{\varepsilon_{33}/\varepsilon_{11}}$ is the dielectric anisotropy factor. Note that while using image charge series is rigorous for linear response, for quadratic responses the set of image charges necessitates calculation of cross-terms in Eq. (2) originating from different charges, and hence is not considered here.

Here we calculate electrostriction response for transverse dielectric isotropy in the absence of the anisotropic part of electrostriction tensor, i.e. under the condition $2Q_{44}^{MT} \equiv Q_{11}^{MT} - Q_{12}^{MT}$. In this case, after integration Eq.(2) acquires the form:

$$u_3^{ES+MT}(0) = \frac{(1+\nu)}{Y}\frac{U^2}{d}\left(Q_{12}^{MT} f_{312} - \left(Q_{11}^{MT} - Q_{12}^{MT}\right) f_{344}\right) \quad (5)$$

where $f_{ijk}$ are universal functions of dielectric anisotropy factor, γ, and Poisson ratio, *v*, as listed in the supplemental material**.** The dependencies of $f_{ijk}$ on the factor γ and *v* are shown in the **Figures 2a,b.** Note that pre-factors $f_{ijk}$ are non-monotonic functions of the anisotropy factor γ, as shown in **Fig. 2 a**. At the same time, **Figure 2b** illustrates *monotonic* dependence of pre-factors $f_{ijk}$ on the Poisson ratio *v* for different values of anisotropy factor γ.

For the case of dielectric isotropic materials with γ=1, Eq. (5) can be simplified as: $u_3^{ES+MT}(0) = g\,U^2/d$ , where

$$g = \frac{(1+\nu)}{Y}\left(Q_{12}^{MT}(1-2\nu)\left(2-\frac{\pi}{2}\right) - \left(Q_{11}^{MT} - Q_{12}^{MT}\right)\left(-\frac{3}{4}+\frac{\pi}{8}+\nu\left(\frac{3}{2}-\frac{3\pi}{8}\right)\right)\right) \quad (6)$$

Below, we evaluate the ES responses for dielectric MgO and paraelectric $SrTiO_3$ based on Eq. (6). Parameters used in the estimations are listed in the **Table 1**. For MgO the coefficient *g* is $g = 4.3\times10^{-22}$ $m^2/V^2$. For $SrTiO_3$ the response is much higher due to the high dielectric susceptibility $g = 8.5\times10^{-20}$ $m^2/V^2$. For applied voltage *U*=10 V and effective tip size *d*=10 nm the electrostrictive responses are $u_3^{ES+MT}(0) = 4.3$ pm for MgO and $u_3^{ES+MT}(0) = 0.85$ nm for $SrTiO_3$.

**Table I.** Parameters used in the estimations of the ES response

| material | $Y = 1/s_{11}$ (GPa) | $\nu = -s_{12}/s_{11}$ | $q_{ij}$ ($10^8$ m V/C) | ε (at RT) | $\varepsilon_\infty$ |
|---|---|---|---|---|---|
| MgO | 249 | 0.239 | $q_{11}$= 850<br>$q_{12}$= 5.4 | 9.7 | 2.9 |
| $SrTiO_3$ | 284 | 0.241 | $q_{11}$= 14<br>$q_{12}$= 16 | 300 | 43 |
| $EuTiO_3$ | 274 | 0.233 | $q_{11}$= 290<br>$q_{12}$= 35 | 156 | 33 |
| $KTaO_3$ | 371 | 0.232 | $q_{11}$= 393<br>$q_{12}$= -17 | 242 | 47.5 |

RT – room temperature

**Figure 2c** illustrates the dependence of the surface displacement (Eq.(6)) induced by the SPM probe vs. the effective dielectric permittivity $\kappa=\sqrt{\varepsilon_{11}\varepsilon_{33}}$ for MgO-, $SrTiO_3$-, $EuTiO_3$- and $KTaO_3$-like materials. Here, "like" describes materials with all parameters similar to that listed in **Table I**, but varying κ. Note that $\kappa\geq\varepsilon_{\infty}$ since the apparent dielectric permittivity is always positive. Note that the response monotonically increases with κ. Similarly, **Figure 2d** illustrates the dependence of the surface displacement vs. electrostriction coupling coefficients ratio $q_{11}/q_{12}$, with the coefficient $q_{12}$ was fixed equal to the known values for MgO, $SrTiO_3$, $EuTiO_3$ and $KTaO_3$. For MgO, the displacement monotonically increases with $q_{11}/q_{12}$, while it changes sign for the $SrTiO_3$, $EuTiO_3$and $KTaO_3$.

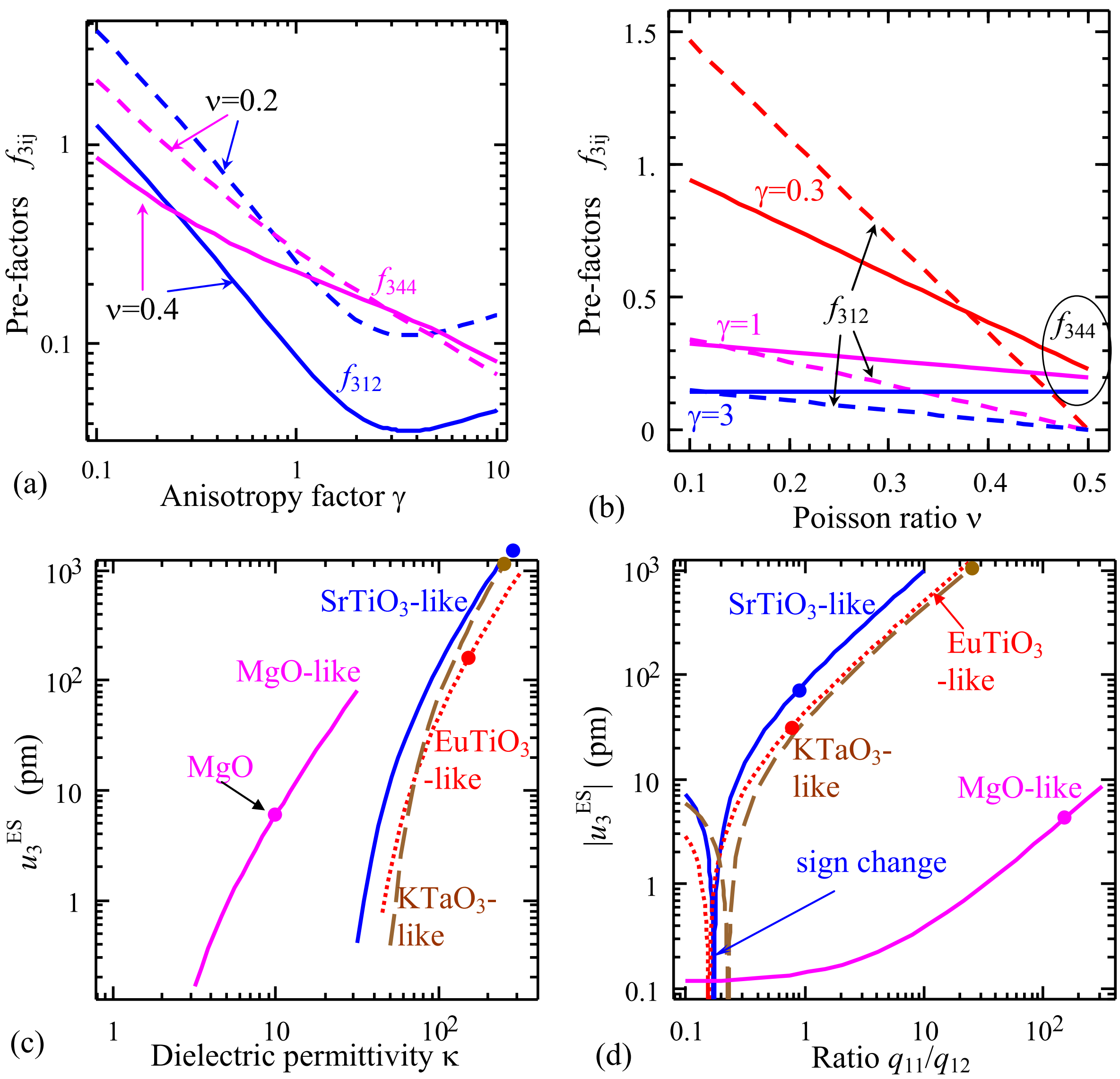

**Figure 2.** The dependence of $f_{312}$ and $f_{344}$ on anisotropy factor **(a)** for different values of Poisson ratio (solid and dashed curves) and on **(b)** on Poisson ratio. **(c)** The dependence of the vertical surface displacement (in pm) on the effective dielectric permittivity κ and **(d)** the ratio of the electrostriction coefficients $q_{11}/q_{12}$ for the dielectrically isotropic materials (γ=1). Wording "material"-like means that all parameters except κ or $q_{11}/q_{12}$ are the same as in **Table I**. For $KTaO_3$ we take $q_{11}/q_{12} > 0$. Point in plots **(c)** and **(d)** corresponds to real materials.

Temperature dependencies of the vertical surface displacement for quantum paraelectrics $SrTiO_3$, $KTaO_3$ and $EuTiO_3$ are shown in the **Figure 3.** For quantum paraelectrics, the temperature dependence of dielectric permittivity is given by Barrett relation $\varepsilon(T)=\varepsilon_\infty+\frac{C_{CW}}{T_q\coth(T_q/T)-T_0}$ [30], where the corresponding parameters are listed in the **Table II**. The electrostriction coefficients are almost temperature independent. Note that $u_3^{ES+MT}(0)$ decreases monotonically with temperature increase. At fixed temperature the displacement is maximal for $SrTiO_3$ with largest permittivity $\varepsilon(T)$, smaller for $KTaO_3$ and minimal for $EuTiO_3$ with the smallest $\varepsilon(T)$.

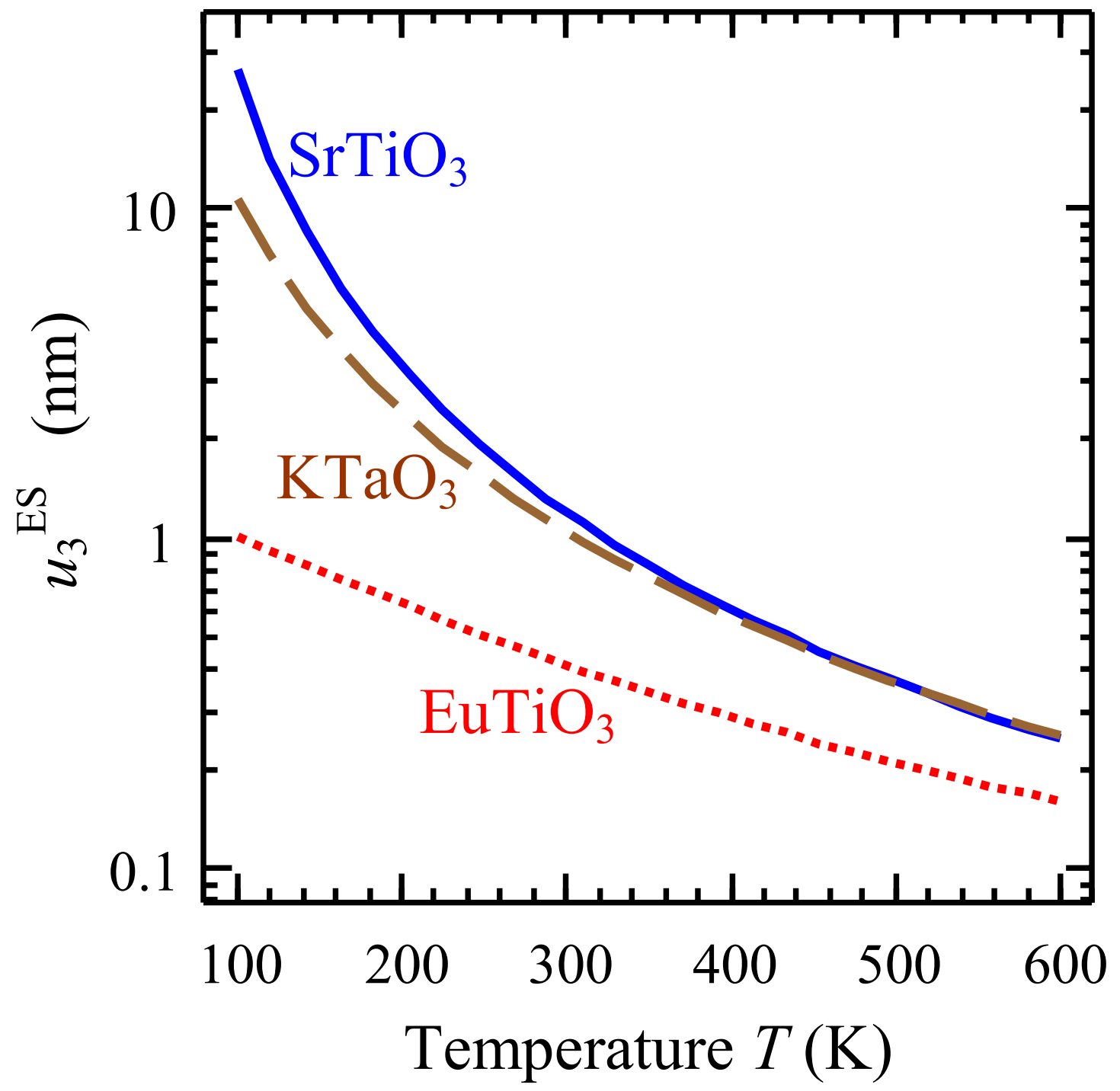

**Figure 3.** Temperature dependencies of the vertical surface displacement for quantum paraelectrics $SrTiO_3$, $KTaO_3$ and $EuTiO_3$

**Table II.** Parameters determining temperature dependent dielectric permittivity of quantum paraelectrics.

| Material | C ($10^5$ K) | $T_0$ (K) | $T_q$ (K) | $\varepsilon_\infty$ | note |
|---|---|---|---|---|---|
| $SrTiO_3$ | 0.507 | 57 | 54 | 43 | Valid at T>105 K |
| $EuTiO_3$ | 0.580 | -165 | 115 | 33 | Valid at T>280 K |
| $KTaO_3$ | 0.545 | 13.1 | 28.5 | 47.5 | |

Finally, we evaluate the electrostatic contribution to PFM/ESM response and compare it with electrostrictive one. First, we consider the tip in ideal contact with the surface and derive associated response. However, this case corresponds to unphysical charge concentration at the tip-surface junction, necessitating the introduction of dielectric tip-surface gap similar to the ferroelectric dead layer. Hence we further discuss the effects of effective gap on dielectric and electrostrictive responses..

The electrostatic displacement of the surface can be evaluated as the electrostatic force $F$ between the effective charge $q$ located at distance $d$ from the surface and its image, divided by the contact stiffness $k$. The force is $F=\frac{1}{4\pi\varepsilon_0\varepsilon_e}\frac{q^2}{d^2}\frac{\varepsilon_e-\kappa}{\varepsilon_e+\kappa}$, $\varepsilon_e$ is the ambient dielectric permittivity, $q$ is the effective charge. The contact stiffness is $k=2R_0Y$, $R_0$ is the contact radius, the effective charge is $q=C_{tip}U$. In a single point charge model $C_{tip}=2\pi\varepsilon_0(\varepsilon_e+\kappa)d$. For the disk-plane case, $C_{tip}=4\varepsilon_0(\varepsilon_e+\kappa)d$ and $d=2R_0/\pi$. Finally, for the sphere-plane case the effective tip capacity is $C_{tip}=4\pi\varepsilon_0\varepsilon_e R_{tip}\frac{\kappa+\varepsilon_e}{\kappa-\varepsilon_e}\ln\left(\frac{\varepsilon_e+\kappa}{2\varepsilon_e}\right)$ and $d\approx 2\varepsilon_e R_{tip}\frac{1}{\kappa-\varepsilon_e}\ln\left(\frac{\varepsilon_e+\kappa}{2\varepsilon_e}\right)$, where $R_{tip}$ is the tip apex curvature. Hence, the electrostatic displacement becomes

$$u_3^{el}=s\frac{\varepsilon_0}{Y\varepsilon_e}(\varepsilon_e-\kappa)(\varepsilon_e+\kappa)\frac{U^2}{R_0} \qquad (7)$$

Where the constant $s=\pi/2$ for the point-charge and sphere-plane cases, and $s=2/\pi$ for the disk-plane case. Remarkably, the electrostrictive and electrostatic contributions scale identically with the tip bias, contact radius, and Young's modulus, and hence cannot be separated in a typical SPM experiment. Rather, the contributing materials constants can be evaluated.

For materials with $\kappa>>1$ and $\varepsilon_e=1$ the prefactor before $U^2/R_0$ is negative and $\frac{\pi\varepsilon_0}{2Y\varepsilon_e}(\varepsilon_e-\kappa)(\varepsilon_e+\kappa)\approx-\frac{\varepsilon_0}{Y}\kappa^2$. In this case, the electrostatic displacement given by Eq.(7) is 10-30 times higher than the electrostriction one given by Eq.(6), as summarized in **Table III**. The ratio $u_3^{el}/u_3^{ES+MT}$ is the highest for the materials with high permittivity κ.

**Table III.** Electrostatic and electrostriction displacements calculated at $U$=10 V, $d$=$R_0$=10 nm, $\varepsilon_e = 1$.

| Material | Electrostatic $u_3^{el}$ (nm) | Electrostriction $u_3^{ES+MT}$ (nm) | Ratio $u_3^{el}/u_3^{ES+MT}$ |
|---|---|---|---|
| MgO | 0.033 | 0.0044 | 7.59 |
| $SrTiO_3$ | 28 | 0.85 | 33.04 |
| $KTaO_3$ | 14 | 1.1 | 12.67 |
| $EuTiO_3$ | 7.8 | 0.42 | 18.38 |

However, we note that the ideal contact case as described here is unphysical. Indeed, the dominant contribution to the electrostatic interaction is caused by the electric field singularity in the contact point, where charge density reaches unphysical values (10 - 50 electrons per unit cell). Small gap between tip and surface can be included into the model to resolve this problem. This approach is similar to dead layer approach[31] which is used in the physics of ferroelectrics to resolve the problem of extremely high magnitudes of the depolarization electric field.

In the case of rigorous sphere-plane model of the tip of curvature $R_{tip}$ located at distance $\Delta R$ from the sample surface (see **Figure 1c**), the image charges are given by recurrent relations $d_{m+1} = R_{tip} + \Delta R - R_{tip}^2 / (R_{tip} + \Delta R + d_m)$ and $Q_{m+1} = Q_m(\kappa - \varepsilon_e) R_{tip} / ((\kappa + \varepsilon_e)(R_{tip} + \Delta R + d_m))$, where $Q_0 = 4\pi\varepsilon_0\varepsilon_e R_{tip} U$, $d_0 = R_{tip} + \Delta R$ and $U$ is tip bias (see e.g., Ref.[32]) The capacitance of the tip is $C_{tip} = \sum_{m=0}^{\infty} Q_m / U$. The electrostatic force acting on the tip is $F = \frac{U^2}{2}\frac{\partial(C_{tip})}{\partial(\Delta R)} = \frac{U}{2}\sum_{m=0}^{\infty}\frac{\partial(Q_m)}{\partial(\Delta R)}$. The displacement induced by the force $F$ can be estimated as $u_3 = F/2R_0 Y$, where the effective contact $R_0 << R_{tip}$.

Ratios of electrostriction to electrostatic displacement $u_3^{ES+MT}/u_3^{el}$ calculated for MgO and $SrTiO_3$ are shown in **Figures 4a** as a function of the gap width, $\Delta R$. Electrostriction contribution dominates for $SrTiO_3$ with a gap > 0.2 nm, i.e. half a unit cell. The ratio is relatively small and almost independent on $\Delta R$ for the case of MgO or $SrTiO_3$ placed in water ambient with $\varepsilon_e = 80$. The ratio monotonically increases with the gap thickness increase in air ($\varepsilon_e = 1$). The increase appeared under the gap thickness increase is much more steep for $SrTiO_3$ than that for MgO.

**Figures 4b-c** show the contour maps of the inverse ratio $u_3^{el}/u_3^{ES+MT}$ calculated for MgO and $SrTiO_3$ in coordinates "ambient permittivity - gap width". Note the difference between these materials, namely almost gap width independent horizontal contours for MgO, and gap width dependent curved contours for $SrTiO_3$. The difference originated from the very different dielectric permittivity of the materials, about 10 for MgO and 300 for $SrTiO_3$.

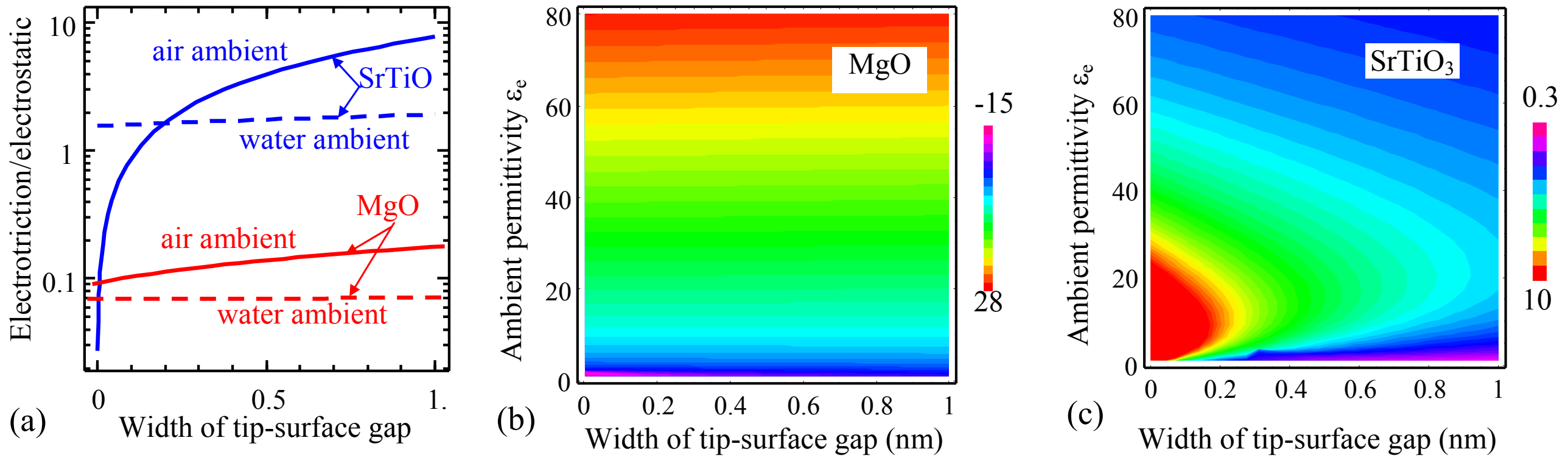

**Figure 4. (a)** Ratios of electrostriction to electrostatic displacement ($u_3^{ES+MT}/u_3^{el}$) calculated for MgO (red curves) and $SrTiO_3$ (blue curves) parameters in dependence on the gap width $\Delta R$ for different ambient permittivity $\varepsilon_e = 1$ (air ambient) and $\varepsilon_e = 80$ (water ambient). Contour maps of the inverse ratio $u_3^{el}/u_3^{ES+MT}$ calculated for MgO **(b)** and $SrTiO_3$ **(c)** in coordinates "ambient permittivity - gap width". The spherical tip radius $R_{tip} = 50$ nm at bias $U$=10 V, $R_0 = 5$ nm.

The vertical electrostatic displacement $u_3^{el}$ as a function of the gap thickness $\Delta R$ between the tip and the surface is shown in **Figures 5a,b**. Note that $u_3^{el}$ is relatively small and almost independent on $\Delta R$ for the case of water ambient. In air $u_3^{el}$ monotonically and rapidly decreases with the gap thickness increase. Dependence of $u_3^{el}$ on the ambient permittivity $\varepsilon_e$ calculated at different $\Delta R = 0.1$, 0.2, 0.3, 0.4, 0.5 and 0.6 nm is shown in **Figures 5c,d**. For $SrTiO_3$ the dependence $u_3^{el}(\varepsilon_e)$ is non-monotonic with a pronounced minimum. The minimum disappears for $\Delta R = 0$ only (see the dashed curve in **Fig.5c**). For MgO the dependence $u_3^{el}(\varepsilon_e)$ is monotonic and almost independent on $\Delta R = 0$ at $\varepsilon_e > 5$. Also note that the response changes its sign at $\varepsilon_e = \varepsilon_{MgO} = 9.7$, because at the value electrostatic forces switch from attractive to repulsive.

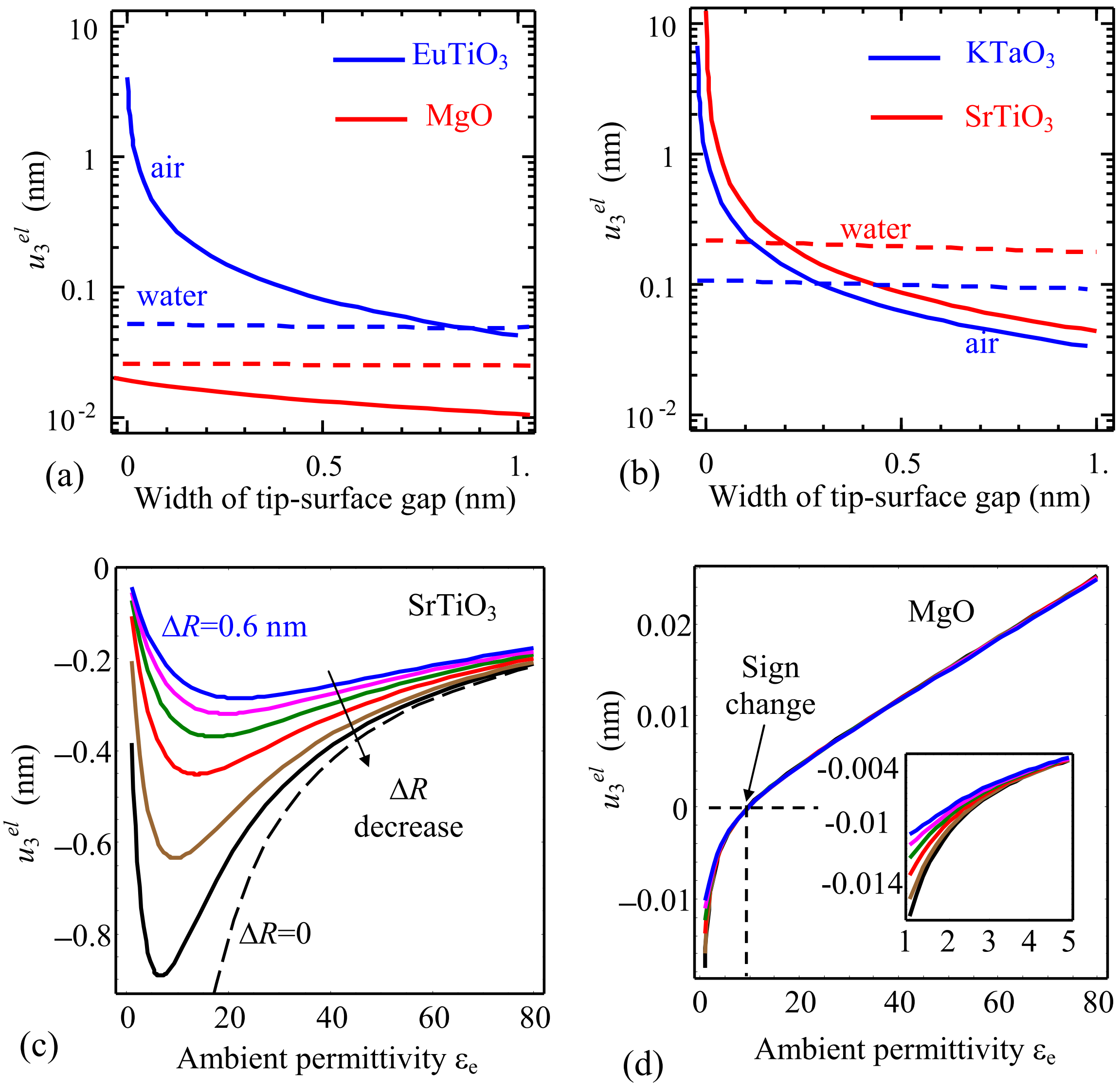

**Figure 5.** Dependence of the electrostatic surface displacement $u_3^{el}$ on the gap $\Delta R$ for different external media, air (solid curves) and water (dashed curves), and for different materials, $EuTiO_3$ and MgO, **(a)** $SrTiO_3$, and $KTaO_3$ **(b)**. Dependence of $u_3^{el}$ on the ambient permittivity $\varepsilon_e$ calculated at $\Delta R = 0$, 0.1, 0.2, 0.3, 0.4, 0.5 and 0.6 nm (different curves) for $SrTiO_3$ **(c)** and MgO **(d)** The spherical tip radius $R_{tip} = 50$ nm at bias $U$=10 V, $R_0 = 5$ nm.

To summarize, the electrostatic and electrostrictive contributions to response signal in contact-mode of voltage modulated SPM have been calculated analytically and numerically. The electrostrictive and electrostatic contributions scale identically with the tip bias, contact radius, and Young's modulus, and hence cannot be separated in a typical SPM experiment. Rather, the contributing materials constants can be evaluated. Calculations showed that electrostatic and electrostriction are strongly dependent on the properties of studied material. For ideal contact, the electrostatic contribution dominates by 1-1.5 orders of magnitude. However, for realistic system maintaining the limited surface charge densities necessitates introduction of effective dielectric gap, similar to ferroelectric dead layers. In the presence of ~1 u.c. thick gap, the electrostrictive contribution dominates for materials such as $SrTiO_3$, and becomes significant for materials with low dielectric constants. The responses also depend strongly on the dielectric constant of gap material. This analysis suggests that rigorous description of bias dependence of PFM and ESM signals necessitates atomistic modelling to evaluate electrostatic responses as controlled by electrostatics of tip-surface junction, whereas developed approximations offer order of magnitude estimates. In comparison, electrostrictive responses can be estimated. Obtained results are important for interpretation of the response signals acquired in piezoresponse force microscopy and electrochemical strain microscopy.

**Acknowledgements:**

A part of this research (AVI, NB, PM, AT, SVK) was conducted at the Center for Nanophase Materials Sciences, which is sponsored at Oak Ridge National Laboratory by the Scientific User Facilities Division, Office of Basic Energy Sciences, U.S. Department of Energy. A.N.M. and E.A.E. are grateful to Prof. S.M. Ryabchenko for multiple discussions and acknowledge the support via bilateral SFFR-NSF project (US National Science Foundation under NSF-DMR-1210588 and State Fund of Fundamental State Fund of Fundamental Research of Ukraine, grant UU48/002).

Supplementary materials for the manuscript

# Electrostrictive and electrostatic responses in contact mode voltage modulated Scanning Probe Microscopies

Anna N. Morozovska[1], Eugene A. Eliseev[2], Anton V. Ievlev[3], Nina Balke[3], Peter Maksymovych[3], Alexander Tselev[3], and Sergei V. Kalinin[3]

[1]Institute of Physics, NAS of Ukraine, 46, pr. Nauki, 03028 Kiev, Ukraine

[2]Institute for Problems of Materials Science, NAS of Ukraine, Krjijanovskogo 3, 03142 Kiev, Ukraine

[3]The Center for Nanophase Materials Sciences, Oak Ridge National Laboratory, Oak Ridge, TN 37922

## Appendix A.

The evident form of the Green function is [F. Felten, G.A. Schneider, J. Muñoz Saldaña, and S.V. Kalinin, J. Appl. Phys. **96**, 563 (2004).]:

$$G_{31}(\boldsymbol{x};\xi)=\frac{(1+\nu)(x_1-\xi_1)}{2\pi Y}\left\{\frac{-\xi_3}{R^3}+\frac{(1-2\nu)}{R(R+\xi_3)}\right\} \tag{A.1a}$$

$$G_{13}(\boldsymbol{x};\xi)=\frac{(1+\nu)(x_1-\xi_1)}{2\pi Y(1-\nu)}\left\{\frac{-\xi_3}{R^3}-\frac{(1-2\nu)}{R(R+\xi_3)}\right\}\neq G_{31}(\boldsymbol{x};\xi) \tag{A.1b}$$

$$G_{32}(\boldsymbol{x};\xi)=\frac{(1+\nu)(x_2-\xi_2)}{2\pi Y}\left\{\frac{-\xi_3}{R^3}+\frac{(1-2\nu)}{R(R+\xi_3)}\right\} \tag{A.1c}$$

$$G_{23}(\boldsymbol{x};\xi)=\frac{(1+\nu)(x_2-\xi_2)}{2\pi Y}\left\{\frac{-\xi_3}{R^3}-\frac{(1-2\nu)}{R(R+\xi_3)}\right\}\neq G_{32}(\boldsymbol{x};\xi) \tag{A.1d}$$

$$G_{33}(\boldsymbol{x};\xi)=\frac{1+\nu}{2\pi Y}\left\{\frac{2(1-\nu)}{R}+\frac{\xi_3{}^2}{R^3}\right\} \tag{A.1e}$$

Here the following designation is introduced $R=\sqrt{(x_1-\xi_1)^2+(x_2-\xi_2)^2+\xi_3^2}$.

Using an approximation of linear dielectrics leads to

$$\frac{\partial G_{ij}^S}{\partial \xi_m} q_{mjkl}^{MT} P_k P_l \approx \frac{q_{1122}^{MT}}{\alpha^2} \frac{\partial G_{ij}^S}{\partial \xi_j} \frac{\partial \varphi}{\partial \xi_k} \frac{\partial \varphi}{\partial \xi_k} + \frac{q_{1111}^{MT} - q_{1122}^{MT}}{\alpha^2} \sum_j \frac{\partial G_{ij}^S}{\partial \xi_j} \left( \frac{\partial \varphi}{\partial \xi_j} \right)^2 + \frac{2 q_{1212}^{MT}}{\alpha^2} \sum_{j \neq k} \frac{\partial G_{ij}^S}{\partial \xi_k} \frac{\partial \varphi}{\partial \xi_k} \frac{\partial \varphi}{\partial \xi_j} \quad \text{(A.2)}$$

Next we suppose that anisotropic part of electrostriction tensor is absent, i.e. $2q_{1212}^{MT} \equiv q_{1111}^{MT} - q_{1122}^{MT}$

$$\frac{\partial G_{ij}^S}{\partial \xi_m} q_{mjkl}^{MT} P_k P_l \approx \frac{q_{1122}^{MT}}{\alpha^2} \left( \frac{\partial G_{ij}^S}{\partial \xi_j} \right) \left( \frac{\partial \varphi}{\partial \xi_k} \frac{\partial \varphi}{\partial \xi_k} \right) + \frac{q_{1111}^{MT} - q_{1122}^{MT}}{\alpha^2} \left( \sum_{k=1}^{3} \sum_{j=1}^{3} \frac{\partial G_{ij}^S}{\partial \xi_k} \frac{\partial \varphi}{\partial \xi_k} \frac{\partial \varphi}{\partial \xi_j} \right) \quad \text{(A.3)}$$

For the case of vertical displacement ($i$=3):

$$\left. \frac{\partial G_{ij}^S}{\partial \xi_j} \frac{\partial \varphi}{\partial \xi_k} \frac{\partial \varphi}{\partial \xi_k} \right|_{\vec{x}=0} =$$

$$= \frac{1+\nu}{2\pi Y} \left( -(1-2\nu) \frac{2\xi_3}{\left( \sqrt{\xi_1^2 + \xi_2^2 + \xi_3^2} \right)^3} \right) \frac{U^2 R_0^2}{\left( \xi_1^2 + \xi_2^2 + (R_0 + \xi_3/\gamma)^2 \right)^2} \left( 1 + \frac{(R_0 + \xi_3/\gamma)^2}{\xi_1^2 + \xi_2^2 + (R_0 + \xi_3/\gamma)^2} \left( \frac{1}{\gamma^2} - 1 \right) \right)$$

$$\sum_{k=1}^{3} \sum_{j=1}^{3} \frac{\partial G_{ij}^S}{\partial \xi_k} \frac{\partial \varphi}{\partial \xi_k} \frac{\partial \varphi}{\partial \xi_j} =$$

$$= \frac{1+\nu}{2\pi Y} \left( \begin{array}{l} (1-2\nu) - \dfrac{4(1-\nu)\xi_3}{\sqrt{\xi_1^2 + \xi_2^2 + \xi_3^2}} - 3 \dfrac{\xi_3^3 \left( \xi_3 - \frac{1}{\gamma} \left( R_0 + \frac{\xi_3}{\gamma} \right) \right)^2}{\left( \sqrt{\xi_1^2 + \xi_2^2 + \xi_3^2} \right)^5} \\ + 2 \dfrac{\xi_3}{\left( \sqrt{\xi_1^2 + \xi_2^2 + \xi_3^2} \right)^3} \left( \xi_3 - \frac{1}{\gamma} \left( R_0 + \frac{\xi_3}{\gamma} \right) \right) \left( (3-\nu)\xi_3 - \frac{\nu}{\gamma} \left( R_0 + \frac{\xi_3}{\gamma} \right) \right) \end{array} \right) \frac{U^2 R_0^2}{\left( \xi_1^2 + \xi_2^2 + (R_0 + \xi_3/\gamma)^2 \right)^3}$$

(A.4)

As a next step we introduce spherical coordinate system as $\xi_1 = \rho \sin\theta \sin\phi$, $\xi_2 = \rho \sin\theta \cos\phi$ and $\xi_3 = \rho\cos\theta$, where $\rho$ is the distance to the coordinate origin, $\theta$ and $\phi$ are the polar and azimuthal angles. Differential element of volume is $d^3\xi = \rho^2 d\rho\, sd\theta\, d\phi$. Integration on angle $\phi$ is reduced to the multiplication on $2\pi$

$$u_{ES}(0) = -2\pi\int_0^\infty \rho^2 dc \int_0^{\pi/2} d\theta \sin\theta \left( \frac{q_{1122}^{MT}}{\alpha^2} \left( \frac{\partial G_{ij}^S}{\partial \xi_j} \right) \left( \frac{\partial \varphi}{\partial \xi_k} \frac{\partial \varphi}{\partial \xi_k} \right) + \frac{q_{1111}^{MT} - q_{1122}^{MT}}{\alpha^2} \left( \sum_{k=1}^{3} \sum_{j=1}^{3} \frac{\partial G_{ij}^S}{\partial \xi_k} \frac{\partial \varphi}{\partial \xi_k} \frac{\partial \varphi}{\partial \xi_j} \right) \right) =$$

$$= \frac{q_{1122}^{MT}}{\alpha^2} \frac{(1+\nu)(1-2\nu)}{Y} \int_0^\infty d\rho \int_0^{\pi/2} d\theta \sin\theta \frac{2\cos\theta U^2 R_0^2}{\left( (\rho\sin\theta)^2 + \left( R_0 + \frac{\rho\cos\theta}{\gamma} \right)^2 \right)^2} \left( 1 + \frac{\left( R_0 + \frac{\rho\cos\theta}{\gamma} \right)^2 \left( \frac{1}{\gamma^2} - 1 \right)}{(\rho\sin\theta)^2 + \left( R_0 + \frac{\rho\cos\theta}{\gamma} \right)^2} \right) -$$

$$- \frac{q_{1111}^{MT} - q_{1122}^{MT}}{\alpha^2} \frac{(1+\nu)}{Y} \int_0^\infty d\rho \int_0^{\pi/2} d\theta \sin\theta \frac{U^2 R_0^2}{\left( (\rho\sin\theta)^2 + \left( R_0 + \frac{\rho\cos\theta}{\gamma} \right)^2 \right)^3} \times$$

$$\left( \begin{array}{l} (1-2\nu)\rho^2 - 4(1-\nu)\rho^2 \cos\theta - 3\cos\theta^3 \left( \rho\cos\theta - \frac{1}{\gamma} \left( R_0 + \frac{\rho\cos\theta}{\gamma} \right) \right)^2 + \\ + 2\cos\theta \left( \rho\cos\theta - \frac{1}{\gamma} \left( R_0 + \frac{\rho\cos\theta}{\gamma} \right) \right) \left( (3-\nu)\rho\cos\theta - \frac{\nu}{\gamma} \left( R_0 + \frac{\rho\cos\theta}{\gamma} \right) \right) \end{array} \right)$$

(A.5)

Here we also suppose that h>>R$_0$. The latter term could be rewritten as

$$- \frac{q_{1111}^{MT} - q_{1122}^{MT}}{\alpha^2} \frac{(1+\nu)}{Y} \int_0^\infty d\rho \int_0^{\pi/2} d\theta \sin\theta \frac{U^2 R_0^2}{\left( (\rho\sin\theta)^2 + \left( R_0 + \frac{\rho\cos\theta}{\gamma} \right)^2 \right)^3} \times$$

$$\left( \begin{array}{l} (1-2\nu)\rho^2 - 4(1-\nu)\rho^2 \cos\theta + \rho^2 \left( 2(3-\nu) - 3(\cos\theta)^2 \right) (\cos\theta)^3 + \\ + \cos\theta \left( 2\nu - 3(\cos\theta)^2 \right) \left( \frac{1}{\gamma} \left( R_0 + \frac{\rho\cos\theta}{\gamma} \right) \right)^2 - 6\rho(\cos\theta)^2 \left( 1 - (\cos\theta)^2 \right) \frac{1}{\gamma} \left( R_0 + \frac{\rho\cos\theta}{\gamma} \right) \end{array} \right) \quad \text{(A.10)}$$

After integration on $\rho$ and $\theta$

$$u_{ES}(0)=\frac{q_{1122}^{MT}}{\alpha^2}\frac{(1+\nu)(1-2\nu)}{Y}\frac{U^2}{R_0}\left(\begin{array}{l}\frac{4\gamma+\pi(\gamma^2-2)+4\sqrt{1-\gamma^2}\arccos(\gamma)}{2\gamma^2}\\ +\left(\frac{1}{\gamma^2}-1\right)\frac{4\gamma+\pi(3\gamma^2-2)+\frac{4(1-2\gamma^2)}{\sqrt{1-\gamma^2}}\arccos(\gamma)}{8\gamma^2}\end{array}\right)-$$

$$-\frac{q_{1111}^{MT}-q_{1122}^{MT}}{\alpha^2}\frac{(1+\nu)}{Y}\frac{U^2}{R_0}\times$$

$$\times\left(\begin{array}{l}(1-2\nu)\frac{\gamma}{12}-(1-\nu)\left(\frac{\gamma}{3}-\frac{4\gamma+\pi(\gamma^2-2)+4\sqrt{1-\gamma^2}\arccos(\gamma)}{4\gamma^2}\right)+\\ \frac{\nu}{\gamma}\left(2-\frac{\pi}{\gamma}\right)+\frac{\gamma}{12}(3-2\nu)-\frac{\pi}{4}(1-\nu)+\frac{\arccos(\gamma)}{2\gamma^2\sqrt{1-\gamma^2}}\left(\gamma^2+\nu(4-3\gamma^2)\right)+\frac{\gamma^2}{4(1-\gamma^2)}\left(\frac{1}{\gamma}-\frac{\arccos(\gamma)}{\sqrt{1-\gamma^2}}\right)-\\ +\frac{\nu}{\gamma^2}\left(\frac{1}{2\gamma}+\frac{\pi(-2+3\gamma^2)}{8\gamma^2}+\left(\frac{1}{2\gamma^2}-1\right)\frac{\arccos(\gamma)}{\sqrt{1-\gamma^2}}\right)+\\ +\frac{1}{\gamma^2}\left(\frac{1}{4(1-\gamma^2)}\left(\gamma-\frac{\arccos(\gamma)}{\sqrt{1-\gamma^2}}\right)-\frac{1}{\gamma}+\frac{\pi(4-3\gamma^2)}{8\gamma^2}-\left(\frac{1}{\gamma^2}-\frac{3}{2}\right)\frac{\arccos(\gamma)}{\sqrt{1-\gamma^2}}\right)-\\ -\left(\frac{1}{2(1-\gamma^2)}\left(\gamma-\frac{\arccos(\gamma)}{\sqrt{1-\gamma^2}}\right)+\frac{1}{\gamma}-\frac{\pi}{2\gamma^2}+\frac{\arccos(\gamma)}{\gamma^2\sqrt{1-\gamma^2}}\right)\end{array}\right)$$

Where coefficient $\alpha=\frac{1}{\varepsilon_0(\kappa-\varepsilon_\infty)}$, effective permittivity $\kappa=\sqrt{\varepsilon_{11}\varepsilon_{33}}$.

## Appendix B

$f_{ijk}$ are analytical functions of dielectric anisotropy $\gamma$

$$f_{312}=(1-2\nu)\left(\left(\frac{1}{2\gamma^2}+1\right)\frac{\sqrt{1-\gamma^2}}{\gamma^2}\arccos(\gamma)+\frac{\pi}{8}+\frac{3}{2\gamma}-\frac{3\pi}{8\gamma^2}+\frac{1}{2\gamma^3}-\frac{\pi}{4\gamma^4}\right), \quad \text{(B.1a)}$$

$$f_{344}=\left(\begin{array}{l}\frac{\arccos(\gamma)}{\sqrt{1-\gamma^2}}\left(-\frac{1}{4}+\frac{5}{4\gamma^2}-\frac{1}{\gamma^4}+\nu\left(\frac{1}{2\gamma^4}-\frac{1}{2}\right)\right)+\\ +\nu\left(\frac{1}{\gamma}+\frac{1}{2\gamma^3}-\frac{\pi}{8\gamma^2}-\frac{\pi}{4\gamma^4}\right)+\frac{1}{4\gamma}-\frac{1}{\gamma^3}-\frac{3\pi}{8\gamma^2}+\frac{\pi}{2\gamma^4}\end{array}\right). \quad \text{(B.1b)}$$

## Appendix C

For the case of transversally-isotropic symmetry of dielectric properties, the potential $V_Q$ in the point charge-based models of the tip has the form:

$$V_Q(\rho,z)=\frac{1}{2\pi\varepsilon_0(\varepsilon_e+\kappa)}\sum_{m=0}^{\infty}\frac{Q_m}{\sqrt{\rho^2+(z/\gamma+d_m)^2}}. \tag{C.1}$$

where $\sqrt{x_1^2+x_2^2}=\rho$ and $\xi_3=z$ are the radial and vertical coordinates respectively, $\varepsilon_e$ is the dielectric constant of the ambient, $\kappa=\sqrt{\varepsilon_{33}\varepsilon_{11}}$ is effective dielectric constant of material, $\gamma=\sqrt{\varepsilon_{33}/\varepsilon_{11}}$ is the dielectric anisotropy factor, $-d_m$ is the z-coordinates of the point charge $Q_m$ and summation is performed over the set of image charges representing the tip.

In the case of rigorous sphere-plane model of the tip of curvature $R_0$ located at distance $\Delta R$ from the sample surface, the image charges are given by recurrent relations $d_{m+1}=R_{tip}+\Delta R-R_{tip}^2\big/\left(R_{tip}+\Delta R+d_m\right)$ and $Q_{m+1}=Q_m(\kappa-\varepsilon_e)R_{tip}\big/\left((\kappa+\varepsilon_e)(R_{tip}+\Delta R+d_m)\right)$, where $Q_0=4\pi\varepsilon_0\varepsilon_e R_{tip}U$, $d_0=R_{tip}+\Delta R$ and $U$ is tip bias.

In the evident form:

$$Q_m=4\pi\varepsilon_0\varepsilon_e R_{tip}U\left(\frac{\varepsilon_i-\varepsilon_e}{\varepsilon_i+\varepsilon_e}\right)^m\frac{\sinh\left(\operatorname{arccosh}\left((R_{tip}+\Delta R)/R_{tip}\right)\right)}{\sinh\left((m+1)\operatorname{arccosh}\left((R_{tip}+\Delta R)/R_{tip}\right)\right)} \tag{C.2a}$$

$$d_m=R_{tip}+\Delta R-\frac{R_{tip}\sinh\left(m\operatorname{arccosh}\left((R_{tip}+\Delta R)/R_{tip}\right)\right)}{\sinh\left((m+1)\operatorname{arccosh}\left((R_{tip}+\Delta R)/R_{tip}\right)\right)} \tag{C.2b}$$

The capacitance of the tip could be written as $C_{tip}=\sum_{m=0}^{\infty}Q_m\Big/U$. The electrostatic force acting on the tip is $F=\frac{U^2}{2}\frac{\partial(C_{tip})}{\partial(\Delta R)}=\frac{U}{2}\sum_{m=0}^{\infty}\frac{\partial(Q_m)}{\partial(\Delta R)}$. Being interested in the limit $\Delta R\to 0$, we derived:

$$\begin{aligned}F_0&=\frac{U^2}{2}\frac{\partial(C_{tip})}{\partial(\Delta R)}\bigg|_{\Delta R\to 0}=-\frac{U^2}{2}4\pi\varepsilon_0\varepsilon_e\sum_{m=0}^{\infty}\frac{m(m+2)}{3(m+1)}\left(\frac{\varepsilon_i-\varepsilon_e}{\varepsilon_i+\varepsilon_e}\right)^m=\\&=-4\pi\varepsilon_0\varepsilon_e\frac{U^2}{2}\frac{1}{3}\left(\left(\frac{\varepsilon_i+\varepsilon_e}{2\varepsilon_e}\right)^2+\frac{\varepsilon_i+\varepsilon_e}{\varepsilon_i-\varepsilon_e}\log\left(\frac{2\varepsilon_e}{\varepsilon_i+\varepsilon_e}\right)\right)\approx-\frac{\pi}{6}\frac{(\varepsilon_i)^2}{\varepsilon_e}\varepsilon_0U^2\end{aligned} \tag{C.2c}$$

Note that the force $F_0$ acting on the sphere near the surface is independent on the sphere radius and grows as $\frac{(\varepsilon_i)^2}{\varepsilon_e}$ with dielectric permittivity $\varepsilon_i$ increase.